\documentclass{article}
\usepackage{array}
\usepackage{longtable}
\usepackage{multirow}
\usepackage{hhline}

\usepackage[dvips]{graphicx}

\usepackage{amsfonts,amssymb, amsmath, amsthm}
\usepackage[english]{babel}

\def\mP{\mathcal{P}}
\def\bbR{\mathbb{R}}
\def\nn{\mathfrak{N}}
\def\ba{\boldsymbol{\alpha}}
\def\bo{\boldsymbol{\omega}}
\def\bO{\boldsymbol{\Omega}}
\def\bb{\boldsymbol{\beta}}

\def\SR{G}
\def\FR{F}
\def\SSR{G^*}
\def\SFR{F^*}

\newcommand \bs[1]{\boldsymbol{#1}}
\newcommand \bss[1]{\mathbf{v}(#1)}
\newcommand {\ts}[1] {\textsl{#1}}
\newcommand{\fns}[1]{{\footnotesize #1}}
\newcommand {\fts}[1] {{\small #1}}
\newcommand {\gs} {\geqslant}

\newcommand{\rul}{\rule[-5pt]{0pt}{18pt}}

\usepackage{hyperref}

\begin{document}

\author{I.I. Kharlamova, A,Yu. Savushkin\\ \\
\it Russian Academy of National Economy and Public Administration\\
\it Volgograd Branch, Volgograd, Russia}

\title{On the geometry of motions in one integrable problem of the rigid body dynamics\footnote{Submitted to J. of Geometry and Physics, 2013}}

\date{}

\maketitle

\begin{abstract}
Due to Poinsot's theorem, the motion of a rigid body about a fixed point is represented as rolling without slipping of the moving hodograph of the angular velocity over the fixed one. If the moving hodograph is a closed curve, visualization of motion is obtained by the method of P.V.\,Kharlamov. For an arbitrary motion in an integrable problem with an axially symmetric force field the moving hodograph densely fills some two-dimensional surface and the fixed one fills a three-dimensional surface. In this paper, we consider the irreducible integrable case in which both hodographs are two-frequency curves. We obtain the equations of bearing surfaces, illustrate the main types of the surfaces. We propose a method of the so-called non-straight geometric interpretation representing the motion of a body as a superposition of two periodic motions.



\end{abstract}

\tableofcontents

\section*{Introduction} 
\begin{figure}[!ht]
\centering
\includegraphics[width=0.3\linewidth,keepaspectratio]{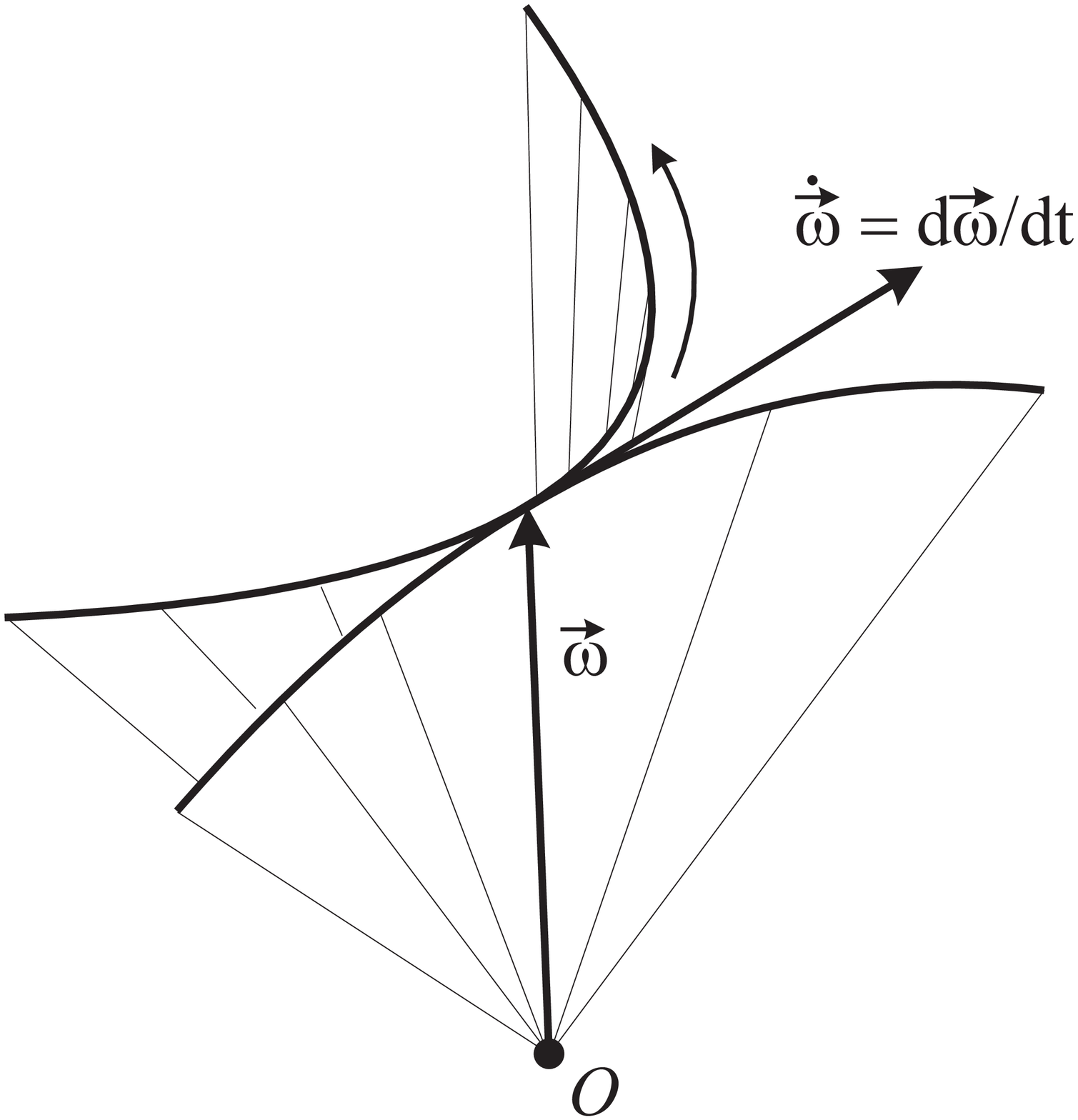}
\caption{Poinsot's theorem.}\label{fig_poin}
\end{figure}

According to the famous result of L.\,Poinsot \cite{Po1851}, an arbitrary motion of a rigid body about a fixed point is represented by rolling without slipping of the moving hodograph of the angular velocity vector over the fixed hodograph of this vector. Since these two curves viewed from the same space have at any time moment the common tangent line, the similar statement is valid also for the conical surfaces generated by the instant rotation axis, namely, the moving axoid is rolling without slipping over the fixed one. The point is that, in the purely rotational motion of a rigid body, the so-called relative and absolute time-derivatives of the angular velocity vector coincide. The general situation is illustrated in Fig.~\ref{fig_poin}. Here $\vec{\omega}$ is the angular velocity, $\dot{\vec{\omega}}$ stands for the relative time-derivative (with respect to some reference frame strictly attached to the body) and $d \vec{\omega}/dt$ is the absolute time-derivative (with respect to an inertial frame).

Estimating his own work, Poinsot wrote that ``it enables us to represent to ourselves the motion of a rigid body as clearly as that of a moving point''. Due to the problems of finding the fixed hodograph, Poinsot gave only one representation of motion, namely, for the Euler case of the body rotation without external forces.

The classical integrable cases in the rigid body dynamics deal with the motions of a rigid body under the influence of axially symmetric potential fields (the homogeneous gravity field, the central Newtonian field). Such systems are called {\it reducible} since the corresponding Euler\,--\,Poisson equations describe the motion up to rotations about the axis of the force symmetry and can be reduced to a Hamiltonian system with two degrees of freedom. This means that while the moving hodograph is completely defined by the solution of the Euler\,--\,Poisson equations, the equations of the fixed hodograph must include an additional quadrature. The same problem arises for the solutions of the Kirchhoff equations of rigid body motion in an ideal fluid, which also can be treated as the equations of a gyrostat motion about a fixed point.

P.V.\,Kharlamov \cite{Kh1964} proposed a natural way to find the fixed hodograph and to investigate its properties for all values of the existing parameters. This method is known as the hodographs method of the kinematic interpretation of motion and is based on applying some non-holonomic kinematic characteristics. If the solution of the Euler\,--\,Poisson equations is periodic, then the moving hodograph is, obviously, a closed curve. In this case the fixed hodograph, as a rule, densely fills a domain on a two-dimensional surface. P.V.\,Kharlamov has shown that this surface is a surface of rotation with the meridian completely defined by the initial periodic solution by means of explicit functions and the missing angular coordinate of the fixed hodograph can be found by integrating the known function of time. The equations obtained in \cite{Kh1964} gave rise to geometric interpretations built for numerous cases of partial integrability (see reviews in  \cite{GorrMono,KhKh1983} and the contemporary state of investigations in \cite{GaGoKov}).

For an arbitrary motion in integrable reducible systems the moving hodograph in the generic case is a two-frequency vector function of time. Therefore, the fixed hodograph for almost all initial data densely fills a three-dimensional region in space. In \cite{Ga1986,Ga1988,Ga1990}, I.N.\,Gashenenko investigated the hodo\-graphs properties for quasi-periodic solutions in the classical cases of Goryachev\,--\,Chaplygin and Kovalevskaya. Using the equations of the paper \cite{Kh1964} and the ideas of the number theory and Fourier analysis, Gashenenko described those classes of non-resonant motions in the integrable reducible rigid body systems for which the components of the angular velocity in the inertial space also are two-periodic functions of time or, in other words, one-valued functions on the corresponding regular Liouville torus (on the connected component of a regular integral manifold of the Euler\,--\,Poisson equations) bearing trajectories with irrational rotation number. The motion is represented by rolling of one surface ``through'' another in such a way that a curve dense in the first surface rolls without slipping over a similar curve dense in the second surface. Still such motions are destroyed by a small perturbation of the integral constants.

In this paper, we consider the case when the force field does not have any symmetry axis. The system then cannot be globally reduced to two degrees of freedom. Nevertheless, in  integrable systems the motions consisting of the points where the first integrals are dependent
play the most important role in the topological analysis of the initial system as a whole, and their geometry can present, clearly enough, the separating cases for different types of the body rotation. Such critical motions are organized into invariant four-dimensional manifolds on which the induced dynamics correspond to Hamiltonian systems with two degrees of freedom called critical subsystems (see e.g. \cite{KhRCD}). Regular integral manifolds of the critical subsystems consist of two-dimensional tori (Liouville tori) and both hodographs belong to the two-dimensional surfaces obtained as images of the corresponding torus under projections from the 6-dimensional phase space onto three-dimensional spaces of the angular velocities (with respect to the rotating body and to some inertial frame). For non-resonant cases the hodographs are dense in these surfaces. The explicit solutions of the critical subsystems allow us to obtain an immediate computer visualization of the surfaces bearing the hodographs. Below we illustrate this process for one of the known systems in the generalized Kovalevskaya case. Simultaneously, we propose another way to describe the body's motion by presenting it as a composition of some simple motions. This composition is based strictly on the known separation of variables.

\section{The explicit solution}\label{sec1} We now consider the system with two degrees of freedom found in \cite{Odin}. Its explicit algebraic solution in separated variables and the rough topological analysis are given in \cite{KhSavUMBeng}. Let us write out the solution in a slightly different form convenient for the purposes of this paper. Suppose that the rigid body with the inertia tensor of the Kovalevskaya type is placed in two linearly independent homogeneous force fields with the centers of application of the fields in the equatorial plane of the body. Let $O$~be the fixed point and $\{\mathbf{e}_1, \mathbf{e}_2, \mathbf{e}_3\}$ the orthonormal basis of the principal inertia axes. The inertia tensor after choosing the dimensionless values becomes $\mathop{\rm diag}\nolimits \{2,2,1\}$. As shown in \cite{KhRCD}, the forces can always be supposed orthogonal and the centers of application can be taken on the principal inertia axes pointed out from $O$ by the vectors $\mathbf{e}_1$ and $\mathbf{e}_2$. Let $\ba$ and $\bb $ be the unit direction vectors of the intensities of the force fields fixed in the inertial space and represented by their components in the moving frame $O\mathbf{e}_1 \mathbf{e}_2 \mathbf{e}_3$. Then the geometric integrals take the form $\bs{\alpha}^2=1$, $\bs{\beta}^2=1$, $\bs{\alpha} \cdot \bs{\beta}=0$, and the total energy of the system (the Hamilton function) is
\begin{equation}\notag
  H = \omega _1^2  + \omega _2^2  + \frac{1}{2}\omega _3^2 - (a\, \alpha_1  + b\, \beta _2 ).
\end{equation}
The scalar characteristics of the force fields without loss of generality can be chosen to satisfy the condition $a > b > 0$. Let us note that in the boundary cases $b=0$ and $b=a$ the problem becomes reducible (the classical Kovalevskaya case and the case of H.M.\,Yehia \cite{Yeh1} with an additional linear integral of a special structure). For arbitrary values of $a$ and $b$ the integrability of the whole irreducible system with three degrees of freedom is proved in \cite{ReySem}. Out of two parameters $a,b$ only one is essential, but we keep both to deal with more symmetric expressions.

The space of the variables $\omega_i, \alpha_j, \beta_k$ is 6-dimensional in virtue of the geometric integrals. In the paper \cite{Odin} two invariant relations are found which define the system considered here. In the sequel, for the sake of brevity we speak about the system $\nn$, meaning the corresponding invariant four-dimensional manifold together with the induced dynamics on it. The system $\nn$ is a critical subsystem \cite{KhRCD} of the general Reyman\,--\,Semenov-Tian-Shansky problem. In addition to the integral $\{H=h\}$ the system $\nn$ has the partial first integral $\{M=m\}$ \cite{Odin}. On $\nn$ the separation of variables is found \cite{KhSavUMBeng}. Let us give the explicit formulas of the solution. The separating variables $s_1,s_2$ are subdue to the general restrictions
\begin{equation}\label{eq1_1}
|s_1| \geqslant a, \qquad |s_2| \leqslant b.
\end{equation}
Let $p=\sqrt{a^2+b^2}>0$ and $ r=\sqrt{a^2-b^2}>0$. Define the constant value $\ell \geqslant 0$ by the following relation
\begin{equation}\label{eq1_2}
  \ell^2 = 2 p^2 m^2 +2 h m +1.
\end{equation}
Denote
\begin{equation}\notag
\begin{array}{c}
\displaystyle{\Psi (s_1 ,s_2 ) = 4 m s_1 s_2  - 2\ell (s_1  + s_2 ) +
\frac{1}{m}(\ell ^2  - 1)},  \qquad \Phi (s)=\Psi(s,s), \\
\displaystyle{\FR _1 = \sqrt {\mathstrut - \Phi (s_1)},} \quad
\displaystyle{\SR _1 = \sqrt {\mathstrut s_1^2 - a^2},} \quad
\displaystyle{\FR _2 = \sqrt {\mathstrut \Phi (s_2)}, } \quad
\displaystyle{\SR _2 = \sqrt {\mathstrut b^2 - s_2^2}.}
\end{array}
\end{equation}
The expressions for the phase variables are
\begin{eqnarray}
& \begin{array}{l}
\displaystyle{\alpha _1  = \frac{\mathstrut 1} {{2 a (s_1  - s_2 )^2
}}[(s_1 s_2 - a^2 )\Psi + \SR _1 \SR _2 \FR  _1 \FR  _2 ], }\\
\displaystyle{\alpha _2  = \frac{\mathstrut 1} {{2a(s_1  - s_2 )^2
}}[(s_1 s_2 - a^2)\FR  _1 \FR  _2  -  \Psi \SR _1 \SR _2], }\\
\displaystyle{\beta _1  =  \frac{\mathstrut 1} {{2b (s_1  - s_2 )^2
}}[\Psi \SR _1 \SR _2 - (s_1 s_2  - b^2)\FR  _1 \FR  _2], }\\
\displaystyle{\beta _2  = \frac{\mathstrut 1} {{2 b (s_1  - s_2 )^2
}}[(s_1 s_2 - b^2 )\Psi + \SR _1 \SR _2 \FR  _1 \FR  _2 ], }\\
\displaystyle{\alpha _3  = \frac{\mathstrut r \, \SR _1 } {a(s_1  - s_2)}
,\quad
\beta _3  =\frac{r \, \SR _2} {b (s_1  - s_2)}, }
\end{array}\label{eq1_3} \\
& \begin{array}{l}
\displaystyle{\omega _1  = \frac{\mathstrut r (\ell - 2ms_1 )\FR  _2 } {{2(s_1  - s_2
)}},\quad \omega _2  = \frac{r (\ell  - 2ms_2)\FR  _1} {{2(s_1 - s_2 )}}, }\quad
\displaystyle{\omega _3  = -\frac{\mathstrut \SR _2
\FR  _1 + \SR _1 \FR  _2 } {{s_1  - s_2 }}}.
\end{array}\label{eq1_4}
\end{eqnarray}
The dependence of $s_1 ,s_2$ on time is given by the completely separated differential equations
\begin{equation}\label{eq1_5}
2 \dot {s}_1 =  \FR _1 \SR _1, \qquad 2 \dot {s}_2 = \FR _2 \SR _2.
\end{equation}

Note that the values $\SR_i,\FR_i$ are supposed algebraic and simultaneous change of the sign of any of these values in \eqref{eq1_3} -- \eqref{eq1_5} does not change the solution as a whole.

\begin{figure}[!ht]
\centering
\includegraphics[width=0.5\linewidth,keepaspectratio]{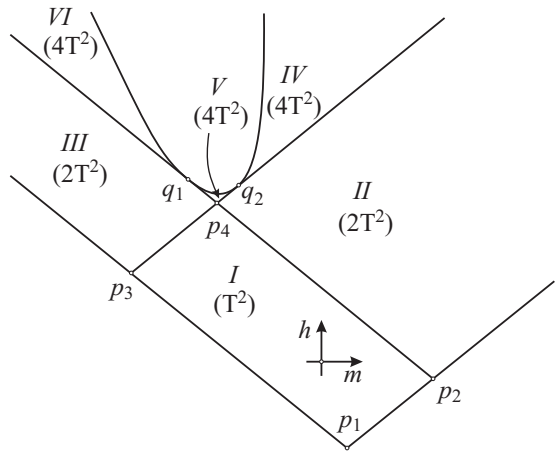}
\caption{The admissible region and the cameras.}\label{fig_bifnum}
\end{figure}

It readily follows from \eqref{eq1_2}, \eqref{eq1_3} and \eqref{eq1_4} that the region of the real motions on the $(m,h)$-plane is defined by the inequalities
\begin{equation}\notag
\begin{array}{c}
h \geqslant \max\{r^2 m -2a, -r^2 m-2b\}, \qquad
2 p^2 m^2+2 h m +1 \geqslant 0.
\end{array}
\end{equation}
This region is usually called the admissible region and any point $(m,h)$ in it is called an admissible point. The bifurcation diagram $\Sigma$ of the first integrals for the solution \eqref{eq1_3} -- \eqref{eq1_5} consists of two half-lines and the part of the hyperbola forming the external boundary of the admissible region
\begin{equation}\notag
\begin{array}{c}
h = r^2 m -2a, \qquad h= -r^2 m-2b \qquad (h \gs -a-b), \\
2 p^2 m^2+2 h m +1 =0  \qquad (m <0),
\end{array}
\end{equation}
and also includes two half-lines inside this region
\begin{equation}\notag
\begin{array}{ll}
  h = r^2 m + 2a & \quad  (h \gs a-b), \\
  h = -r^2 m + 2b & \quad  (h \gs -a+b).
\end{array}
\end{equation}
All half-lines correspond to multiple roots of the polynomials $\SR_1^2 \FR_1^2$ and $\SR_2^2 \FR_2^2$. During such motion one of the separating variables remains constant.

In the connected component of $\bbR^2(m,h)\backslash \Sigma$ the topology of regular integral manifolds does not change. In Fig.~\ref{fig_bifnum} we enumerate cameras in the admissible region and show the number of Liouville tori in the integral manifold according to the results of the work \cite{KhSavUMBeng}. The points $p_1,\ldots,p_4$ are the images of four existing equilibria. Though the hyperbola is a part of the external boundary of the admissible region, the integral manifolds in its pre-image $\mathcal{L}_0=\{\ell=0\}\subset \nn$ consist of two-dimensional tori except for two points $q_1, q_2$ when the bifurcations take place inside the three-dimensional submanifold $\mathcal{L}_0$.

Writing out the solutions of equations \eqref{eq1_5} in Jacobi functions, we obtain, as the arguments of these functions, the angular coordinates on the Liouville tori. The corresponding expressions for all cases can be found in \cite{SavPHD}.


\section{Decomposition of motions}\label{sec2}
Let us introduce some terminology compatible with that of \cite{Ga1986,Ga1990}. In addition to the moving basis, we choose an orthonormal basis $\{\mathbf{n}_1, \mathbf{n}_2, \mathbf{n}_3\}$ fixed in the inertial space and call $O\mathbf{n}_1 \mathbf{n}_2 \mathbf{n}_3$ the fixed frame.
To each physical vector of the angular velocity $\vec \omega = \omega_i \mathbf{e}_i = \Omega_i \mathbf{n}_i$ we assign two representatives in $\bbR^3$
\begin{equation}\notag
\bo = (\omega_1,\omega_2,\omega_3) \quad  \mathrm{and} \quad \bO = (\Omega_1, \Omega_2, \Omega_3).
\end{equation}
The cartesian spaces $\{\bo\}$ and $\{\bO\}$ are called the $P$-space and the $\Pi$-space respectively. Let us consider an integral manifold fixing the values of all first integrals. The image of a connected component of this manifold in the $P$-space (the $\Pi$-space) is called a $P$-surface \cite{Ga1986} (respectively, a $\Pi$-surface \cite{Ga1990}). For brevity, both surfaces are called bearing surfaces.

The body's orientation in space is defined by the orientation matrix $Q$ (also called the matrix of direction cosines) such that
\begin{equation}\label{eq2_1}
  \mathbf{n}_{i} = Q_{ik} \mathbf{e}_{k}.
\end{equation}
The vectors $\bo$ and $\bO$ are then connected by the relation
\begin{equation}\label{eq2_2}
  \bO  = Q \bo.
\end{equation}

The set of all orientation matrices is the group $SO(3)$. The images of connected components of  integral manifolds in $SO(3)$ will be called $Q$-surfaces. A smooth curve in a $Q$-surface will be called a \textit{possible} motion.

Let $\mathbf{v}: \bbR^3 \to so(3)$ be the isomorphism of the algebras
\begin{equation}\notag
  \mathbf{a} = (a_k) \mapsto A  = (a_{ij}), \qquad a_{ij}=-\varepsilon_{ijk} a_k.
\end{equation}
On any motion $Q(t)$ the corresponding vectors $\bo$ and $\bO$ are calculated by the rules
\begin{equation}\label{eq2_3}
  \bo = \bss{ Q^T \dot Q}, \qquad \bO = \bss{ \dot Q Q^T}.
\end{equation}
Any continuous motion $Q(t)$ can be physically realized but the pair $(Q(t),\bo(t))$ defined by \eqref{eq2_3} will not generally satisfy the dynamical equations of motion. Even if the motion $Q(t)$ is possible, the pair $(Q(t),\bo(t))$ can be not compatible with the integral relations. If this pair satisfies the equations of motion and the chosen integral relations, then we call $Q(t)$ the \textit{real} motion.


Suppose we deal with a partial integrable problem of rigid body motion with a four-dimensional phase space $\mP \subset \{(Q,\bo)\}$ and let $J(\mathbf{c})$ be an integral manifold defined by any complete set of the integral constants $\mathbf{c}$. By the Liouville\,--\,Arnold theorem this manifold consists of two-dimensional tori with quasi-periodic trajectories. If the problem is reducible (the Hamilton function admits a symmetry group of rotations about an axis fixed in the inertial space), then the $P$-surface appears to be one-dimensional, i.e., the moving hodograph is a closed curve. In irreducible systems the $P$-, $\Pi$- and $Q$-surfaces of the manifold $J(\mathbf{c})$ are two-dimensional. More precisely, these surfaces are the images of the 2-torus under smooth mappings. It means that each phase variable is a function of two angular coordinates on the torus and these coordinates are linear functions of time. Naturally, we consider a non-degenerate system, i.e., non-resonant tori are dense in $\mP$. In this case we say that all phase variables are two-periodic functions of time. In fact, to obtain such expressions one needs to have an explicit separation of variables. Suppose that on real motions the orientation matrix $Q$ is a function of two auxiliary variables $x_1,x_2$ and the integral constants $\mathbf{c}$, and the equations of motion separate
\begin{equation*}
  \frac{dx_i}{d\tau}=\sqrt{f_i(x_i;\mathbf{c})} \qquad (i=1,2),
\end{equation*}
where $\tau$ is a monotonous function of $t$ defined by a finite or a differential equation. In our system $\nn$ equations \eqref{eq1_5} are completely separated and $\tau=t$. In the regular two-periodic case all $f_i$ have only simple roots and the variables $x_i$ as functions of the ``reduced'' time $\tau$ independently oscillate between two adjacent roots of the corresponding $f_i$. On each Liouville torus in $J(\mathbf{c})$, the separated equations allow us to introduce angular coordinates $(\psi_1,\psi_2 \mathop{\rm modd}\nolimits 2\pi)$ with constant frequencies in the ``reduced'' time
\begin{equation}\notag
  \psi_i = \psi^{0}_i+ C_i(\mathbf{c}) \tau.
\end{equation}
On the $(x_1,x_2)$-plane the motion is represented as a ``sum'' of two independent \textit{partial} motions $(x_1(\psi_1(\tau)),x_2^{0})$ and $(x_1^{0},x_2(\psi_2(\tau)))$. For the orientation matrix and the angular velocity vectors we obtain according to \eqref{eq2_3}
\begin{equation}\notag
\begin{array}{l}
  \bo = \bo_{(1)} + \bo_{(2)}, \qquad \bO = \bO_{(1)} + \bO_{(2)}, \\[2mm]
  \displaystyle \bo_{(i)} =\bss {Q^T(x_1,x_2) \frac{\partial Q(x_1,x_2)}{\partial x_i} } \frac{d x_i}{d\tau} \dot \tau, \\
  \displaystyle \bO_{(i)} =\bss {\frac{\partial Q(x_1,x_2)}{\partial x_i} Q^T(x_1,x_2) } \frac{d x_i}{d\tau} \dot \tau,
\end{array}
\end{equation}
i.e., the angular velocity vectors of a two-periodic motion really are the sums of the corresponding angular velocities of the partial motions. Note that the partial motions are possible motions. If they are simple enough, their properties can give a clear representation of the motion as a whole.

Another way to decompose two-periodic motions is dealing with some intermediate frames of references. Such a frame is sometimes called semi-moving or semi-fixed and is defined by a trihedral which is rotating in quite a simple and clear way with respect to the moving or to the fixed frame. Suppose, for example, that we can write
\begin{equation}\label{eq2_4}
  Q(x_1,x_2)=Q''(x_1,x_2)Q'(x_1,x_2), \qquad Q',Q'' \in SO(3),
\end{equation}
where $Q'(x_1,x_2)$ has a simple form. Then the matrix $Q''$ describes the orientation in the inertial space of the trihedral $O\mathbf{n}'_1 \mathbf{n}'_2 \mathbf{n}'_3$ and with respect to it the rotation of the moving frame $O\mathbf{e}_1 \mathbf{e}_2 \mathbf{e}_3$ is a ``simple'' motion with the matrix $Q'$:
\begin{equation}\notag
  \mathbf{n}'_{i} = Q'_{ik} \mathbf{e}_{k}.
\end{equation}
Putting
\begin{equation}\notag
  \bo' = \bss {Q'^T \dot Q'}, \qquad \bo'' = \bss{Q''^T \dot Q''},
\end{equation}
we get the decomposition of the angular velocity in the moving basis
\begin{equation}\notag
  \bo = \bo'+Q'^T \bo''.
\end{equation}
Similar decomposition can be obtained for the vector $\bO$ if the matrix $Q''$ has a simple expression and the semi-moving frame is in a simple motion with respect to the fixed frame.

Let us return to the investigated system $\nn$. The specific feature of exact solutions in irreducible systems is that the solution itself contains the expression for the orientation matrix $Q$. Indeed, by definition \eqref{eq2_1}
\begin{equation}\notag
  Q=\begin{pmatrix} \alpha_1 & \alpha_2 & \alpha_3 \\
  \beta_1 & \beta_2 & \beta_3   \\
  \gamma_1 & \gamma_2 & \gamma_3
   \end{pmatrix},
\end{equation}
where the first two rows are given by expressions \eqref{eq1_3} and for the components of the vector $\bs{\gamma} = \ba{\times}\bb$ we have
\begin{equation}\label{eq2_5}
\begin{array}{c}
\displaystyle \gamma _1  = -\frac{\mathstrut r (s_2 \SR _1 \Psi  + s_1 \SR _2 \FR  _1 \FR  _2 )} {{2 a b (s_1  - s_2 )^2}}, \quad
\displaystyle \gamma _2  = -\frac{\mathstrut r(s_2 \SR _1 \FR  _1 \FR  _2  - s_1 \SR _2  \Psi)} {{2 a b (s_1  - s_2 )^2}}, \\
\displaystyle \gamma _3  = -\frac{\mathstrut a^2 s_2 - b^2 s_1} {{2 a b (s_1  - s_2 )}}.
\end{array}
\end{equation}
The obtained expressions \eqref{eq1_3}, \eqref{eq2_5} give rise to the decomposition of the type \eqref{eq2_4} with the matrices
\begin{eqnarray}
&  \displaystyle Q'=\frac{1}{2(s_1-s_2)} \begin{pmatrix} -\Psi           & - \FR _1 \FR _2 & 0 \\
                                             \FR _1 \FR _2 & - \Psi          & 0 \\
                                             0             & 0               & 2(s_1-s_2) \end{pmatrix},\nonumber \\
&  \displaystyle Q''=\frac{1}{a b (s_1-s_2)} \begin{pmatrix} b(a^2-s_1 s_2)  & b \, \SR _1 \SR _2  & b\,r \SR _1 \\
                                              -a \, \SR _1 \SR _2 & a(b^2-s_1 s_2)   & a\,r \SR _2 \\  r \SR _1 s_2     & -r \SR _2 s_1    & b^2 s_1-a^2 s_2 \end{pmatrix}. \nonumber
\end{eqnarray}
Obviously, with respect to the body the motion of the semi-moving basis is of pendulum type with the axis of the pendulum coinciding with the dynamical symmetry axis  $O\mathbf{e}_3$. The corresponding angular velocity is
\begin{equation}\label{eq2_6}
  \bo' = \bss {Q'^T \dot Q'} = \left(0,0,-\frac{\FR _1 \SR _2 + \FR_2 \SR _1}{2(s_1-s_2)}\right).
\end{equation}
Here as usual we say that the motion is of pendulum type if it is periodic with a fixed axis of instant rotation. In what follows, we call a pendulum type motion an \textit{oscillation} if the angular velocity periodically changes its direction to the opposite one. If the direction of the angular velocity is constant we say that a pendulum type motion is a \textit{rotation}.

In accordance with the separation of variables \eqref{eq1_5}, the partial motions of the semi-moving basis have the angular velocities
\begin{equation}\label{eq2_7}
\begin{array}{l}
  \displaystyle \bo'_{(1)} =\bss{ Q'^T \frac{\partial Q'}{\partial s_1} } \dot s_1 =
  -\frac{\SR _1}{2(s_1-s_2)} \left(  0, 0, \FR _2  \right), \\
  \displaystyle \bo'_{(2)} =\bss {Q'^T \frac{\partial Q'}{\partial s_2} } \dot s_2 =
  -\frac{\SR _2}{2(s_1-s_2)} \left(  0, 0, \FR _1  \right).
\end{array}
\end{equation}
In the first partial motion $s_2={\rm const}$ and, consequently, $\FR _2={\rm const}$. For the second partial motion we have, respectively, $s_1={\rm const}$ and $\FR _1={\rm const}$. Therefore, the character of the pendulum motions \eqref{eq2_7} is defined by the scalar multiple in the right-hand sided.

The motion $Q''$ is more complicated. Let us calculate for it the angular velocities of the partial motions:
\begin{equation}\label{eq2_8}
\begin{array}{l}
  \displaystyle \bo''_{(1)} =\bss {Q''^T \frac{\partial Q''}{\partial s_1} } \dot s_1 =
  -\frac{\FR _1}{2(s_1-s_2)} \left(  0, -r, \SR _2  \right), \\
  \displaystyle \bo''_{(2)} =\bss {Q''^T \frac{\partial Q''}{\partial s_2} } \dot s_2 =
  -\frac{\FR _2}{2(s_1-s_2)} \left(  r , 0, \SR _1  \right).
\end{array}
\end{equation}
Since in the first partial motion $\SR _2={\rm const}$, it is a pendulum type motion with the axis in the plane $O\mathbf{e}_2\mathbf{e}_3$. In the second partial motion $\SR _1={\rm const}$. Then it is also a pendulum type motion but with the axis in the plane $O\mathbf{e}_1\mathbf{e}_3$. More detailed statements on these motions will be obtained from the segments of oscillation of the separating variables and the corresponding evolution of the radicals $\SR_i,\FR_i$.

\section{Bearing surfaces and partial motions}\label{sec4}
Using equations \eqref{eq1_3} and \eqref{eq2_5} for the orientation matrix, we easily find from \eqref{eq1_4}, \eqref{eq2_2} the components of the angular velocity in the fixed frame:

\begin{eqnarray}
& & \displaystyle \Omega _1  =  \displaystyle - \frac{\mathstrut r} {{2 a (s_1  - s_2 )^2}}
[\SR _1 \SR  _2 \FR  _1 -(a^2+s_1 s_2 -2 s_1^2)\FR _2  ], \nonumber \\
& & \displaystyle \Omega _2  =  \displaystyle -\frac{\mathstrut r} {{2 b (s_1  - s_2 )^2}}
[\SR _1 \SR  _2 \FR  _2 + (b^2+s_1 s_2 -2 s_2^2)\FR _1], \label{eq3_1}\\
& & \displaystyle \Omega _3  =  \displaystyle - \frac{\mathstrut 1} {{2 a b (s_1  - s_2 )^2}}
\{ [b^2 s_1 + a^2(s_1-2 s_2)]\FR _1 \SR _2  - \nonumber \\[2mm]
& & {\phantom{\Omega _3  = }}  \displaystyle - [a^2 s_2 - b^2(2s_1-s_2)]\FR _2 \SR _1 \}.\nonumber \end{eqnarray}

Equations \eqref{eq1_4} and \eqref{eq3_1} provide an analytical foundation for the computer visualization of the $P$- and $\Pi$-surfaces. Nevertheless, it is necessary to make some remarks.

It is clear from \eqref{eq1_3} -- \eqref{eq1_5} that the variables $s_1,s_2$ oscillate in the segments defined by the inequalities $\SR ^2 _i (s_i) \gs 0$, $\FR ^2 _i (s_i) \gs 0$ ($i=1,2$), which in addition to the restrictions \eqref{eq1_1} give
\begin{equation}\notag
  - m (s_1-\theta_1)(s_1-\theta_2) \gs 0, \qquad m (s_2-\theta_1)(s_2-\theta_2) \gs 0.
\end{equation}
Here
\begin{equation}\notag
\displaystyle \theta_1 = \frac{\ell-1}{2m}, \qquad \theta_2 = \frac{\ell+1}{2m}
\end{equation}
are the roots of $\Phi(s)$.

The segments of oscillation are usually called the accessible regions. They depend on the integral constants. If the integral constants are fixed and both $s_1$ and $s_2$ are chosen strictly inside their accessible regions, the arbitrary signs of $\SR_i,\FR_i$ define 16 points on the integral manifold projecting to the point $(s_1,s_2)$. Some of them belong to the same connected component. If, for example, the accessible region for $s_i$ is $[\lambda,\mu]$, then each time the variable reaches the end of the segment the corresponding algebraic value $\sqrt{\mu-s_i}$ or $\sqrt{s_i-\lambda}$ changes its sign. The radicals which are not involved in the definition of the accessible region keep constant sign along the considered trajectory. The choice of these constant signs affects the choice of the connected component of the integral manifold (see \cite{KhND11}). Fixing these signs and making obvious trigonometric substitutions we express the components of $\bo$ and $\bO$ given by \eqref{eq1_4} and \eqref{eq3_1} as one-valued functions of some angular coordinates on the chosen Liouville torus. This gives us an explicit parametrization of the bearing surfaces.

Still there is another problem specific for the investigated system $\nn$. While for $s_2$ the accessible region is always contained in the bounded segment $[-b,b]$, the variable $s_1$ for $m<0$ periodically crosses the infinity. Indeed, in this case the right-hand part of the first equation \eqref{eq1_5} is the square root of the polynomial of degree 4 with positive principal coefficient. Formally speaking, all corresponding singularities in the expressions for the phase variables are removable and the sign of $m$ has no influence on the phase topology. But from the point of view of computation algorithms, to obtain continuous functions we need, each time when $s_1$ crosses the infinity, to change the signs of $\SR_1$ and $\FR_1$. In \cite{KhND11}, for the purpose of the topological analysis this problem is solved by introducing the new values
\begin{equation}\label{eq3_2}
  \SSR_1=s_1^{-1}\SR_1, \qquad \SFR _1=s_1^{-1}\FR_1.
\end{equation}
They do not change signs at $s_1=\infty$ since $s_1$ jumps from $+\infty$ to $-\infty$. Here we use the same approach and make trigonometric substitutions for the variable $s_1^{-1}$, which oscillates in a bounded region.

The complete information on the accessible regions is collected in Table~1. The $(\pm)$ sign attached to the notation of some cameras show the part of the camera with corresponding sign of $m$ (see Fig.~\ref{fig_bifnum}). For negative values of $m$ we write $s_1\in [\,A(\pm
\infty)B\,]$ meaning that $B<0<A$ and $s_1^{-1}$ oscillates on the segment $[\, B^{-1}, A^{-1}\,]$.

\begin{longtable}{|c| c| c| c|}
\multicolumn{4}{r}{\fts{Table 1}}\\
\hline
{\renewcommand{\arraystretch}{0.8}\fns{\begin{tabular}{c}Camera\end{tabular}}}&{\renewcommand{\arraystretch}{0.8}\fns{\begin{tabular}{c}Roots of $\Phi$\end{tabular}}}&{\renewcommand{\arraystretch}{0.8}\fns{\begin{tabular}{c}Oscillation\\segment of $s_1$\end{tabular}}}
&{\renewcommand{\arraystretch}{0.8}\fns{\begin{tabular}{c}Oscillation\\segment of $s_2$\end{tabular}}}\\
\hline\endfirsthead %
\multicolumn{4}{r}{\fts{Table 1 (continued)}}\\
\hline
{\renewcommand{\arraystretch}{0.8}\fns{\begin{tabular}{c}Camera\end{tabular}}}&{\renewcommand{\arraystretch}{0.8}\fns{\begin{tabular}{c}Roots of $\Phi$\end{tabular}}}&{\renewcommand{\arraystretch}{0.8}\fns{\begin{tabular}{c}Oscillation\\segment of $s_1$\end{tabular}}}
&{\renewcommand{\arraystretch}{0.8}\fns{\begin{tabular}{c}Oscillation\\segment of $s_2$\end{tabular}}}\\
\hline\endhead
\rul $\ts{I}_+$&$ - b < \theta _1  < b < a < \theta _2 $&$[\,a,\theta _2 \,]$&$[\, - b,\theta _1 \,]$\\
\hline
\rul $\ts{I}_-$&$\theta _2  <  - a <  - b < \theta _1  < b < a$&$[\,a \,(\pm \infty)\, \theta _2 \,]$&$[\, - b,\theta _1 \,]$\\
\hline
\rul $\ts{II}_+$&$b < \theta _1  < a < \theta _2 $&$[\,a,\theta _2 \,]$&$[\, - b,b\,]$\\
\hline
\rul $\ts{II}_-$&$\theta _2  <  - a <  - b < b < \theta _1  < a$&$[\,a \,(\pm \infty)\, \theta _2 \,]$&$[\, - b,b\,]$\\
\hline
\rul $\ts{III}$&$ - a < \theta _2  <  - b < \theta _1  < b < a$&$[\,a \,(\pm \infty)\, -a\,]$&$[\, - b,\theta _1 \,]$\\
\hline
\rul $\ts{IV}_+$&$a < \theta _1  < \theta _2 $&$[\,\theta _1 ,\theta _2 \,]$&$[\, - b,b\,]$\\
\hline
\rul $\ts{IV}_-$&$\theta _2  <  - a < a < \theta _1 $&$[\,\theta _1 \,(\pm \infty)\, \theta _2 \,]$&$[\, - b,b\,]$\\
\hline
\rul $\ts{V}$&$ - a < \theta _2  <  - b < b < \theta _1  < a$&$[\,a \,(\pm \infty)\, -a\,]$&$[\, - b,b\,]$\\
\hline
\rul $\ts{VI}$&$ - a <  - b < \theta _2  < \theta _1  < b < a$&$[\,a \,(\pm \infty)\, -a\,]$&$[\,\theta _2 ,\theta _1 \,]$\\
\hline
\end{longtable}


Note that the image of the integral manifold both in $P$-space and in $\Pi$-space (including all connected component) is symmetric with respect to all three coordinate planes. Indeed, to change the signs of $\omega_1$ and $\Omega_1$ one needs to simultaneously change the signs of the radicals $\SR_1,\FR_2$. Other components of $\bo$ and $\bO$ stay unchanged. In the same way, changing the signs of $\SR_2,\FR_1$ we change the signs only of $\omega_2,\Omega_2$ and
changing the signs of $\SR_1,\SR_2$ we change the signs only of $\omega_3,\Omega_3$.
For the connected $P$-surfaces and $\Pi$-surfaces, we can see from Table~1 that they are symmetric with respect to the planes $\omega_3=0$ and $\Omega_3=0$ in the cameras $\ts{I},\ts{II},\ts{III}$ and $\ts{V}$, to the planes $\omega_2=0$ and $\Omega_2=0$ in the camera $\ts{IV}$ and
to the planes $\omega_1=0$ and $\Omega_1=0$ in the camera $\ts{VI}$.
Therefore, if for given values of the integral constants we have several surfaces (namely, two or four as in Fig.~\ref{fig_bifnum}), then all of them can be obtained from one surface by an appropriate reflection with respect to some coordinate plane. Thus, to visualize $P$- and $\Pi$-surfaces it is enough to restrict ourselves to only one choice of the signs for those radicals out of $\SR_i,\FR_i$ which have constant sign on each surface.

\def\TableTwoCof{0.33}

\begin{longtable}{|c| c| c|}
\multicolumn{3}{r}{\fts{Table 2}}\\
\hline
{\renewcommand{\arraystretch}{0.8}\fns{\begin{tabular}{c} Camera\end{tabular}}}
&
\fns{\begin{tabular}{c}$P$-surface\end{tabular}}
&
\fns{\begin{tabular}{c}$\Pi$-surface\end{tabular}}
\\
\hline\endfirsthead %
\multicolumn{3}{r}{\fts{Table 2 (contitinued)}}\\
\hline
{\renewcommand{\arraystretch}{0.8}\fns{\begin{tabular}{c} Camera\end{tabular}}}
&
\fns{\begin{tabular}{c}$P$-surface\end{tabular}}
&
\fns{\begin{tabular}{c}$\Pi$-surface\end{tabular}}
\\
\hline\endhead
\begin{tabular}{c}\ts{I}\end{tabular} & \begin{tabular}{c}\includegraphics[width=\TableTwoCof\linewidth,keepaspectratio]{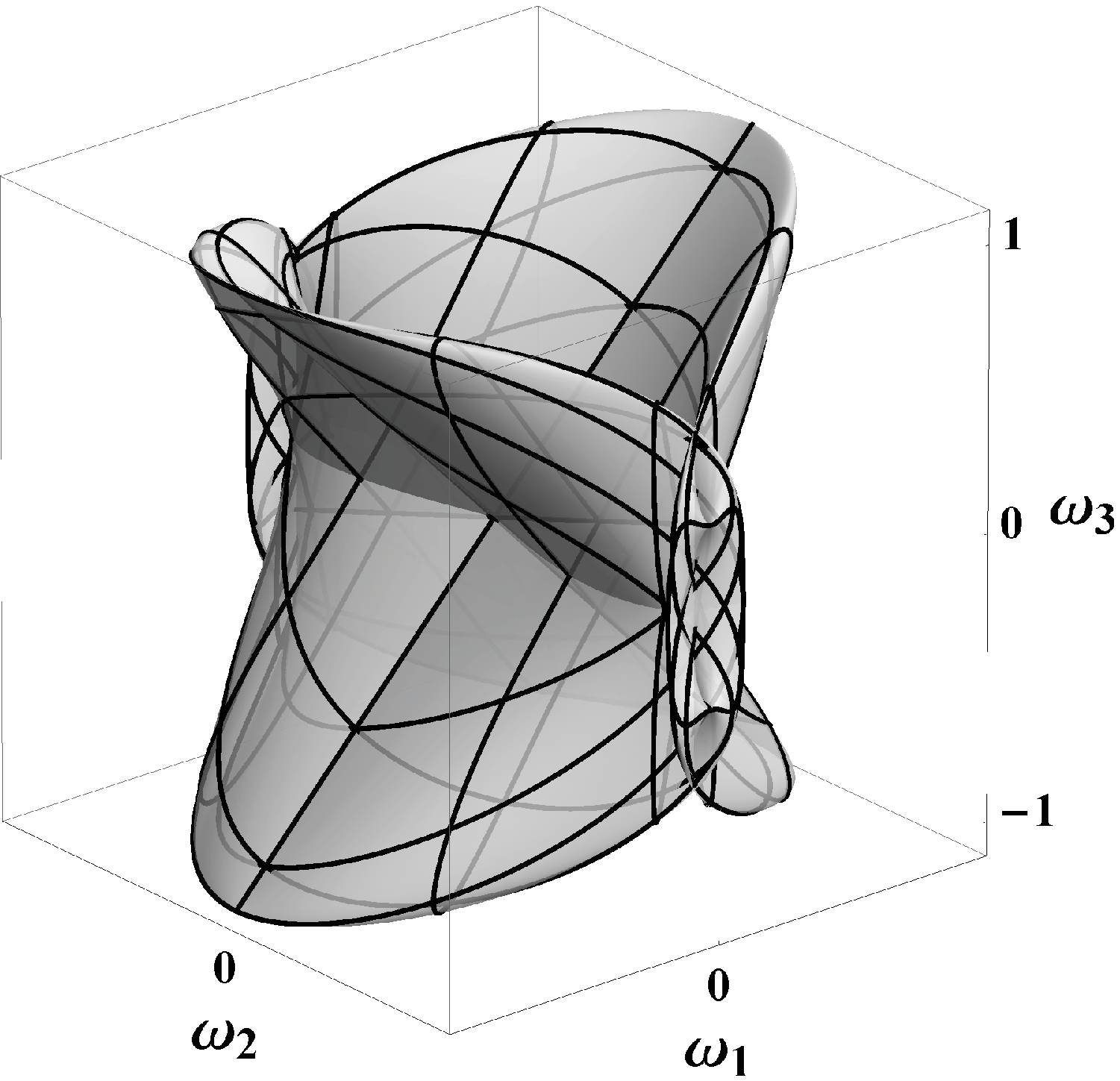}\end{tabular} & \begin{tabular}{c}\includegraphics[width=\TableTwoCof\linewidth,keepaspectratio]{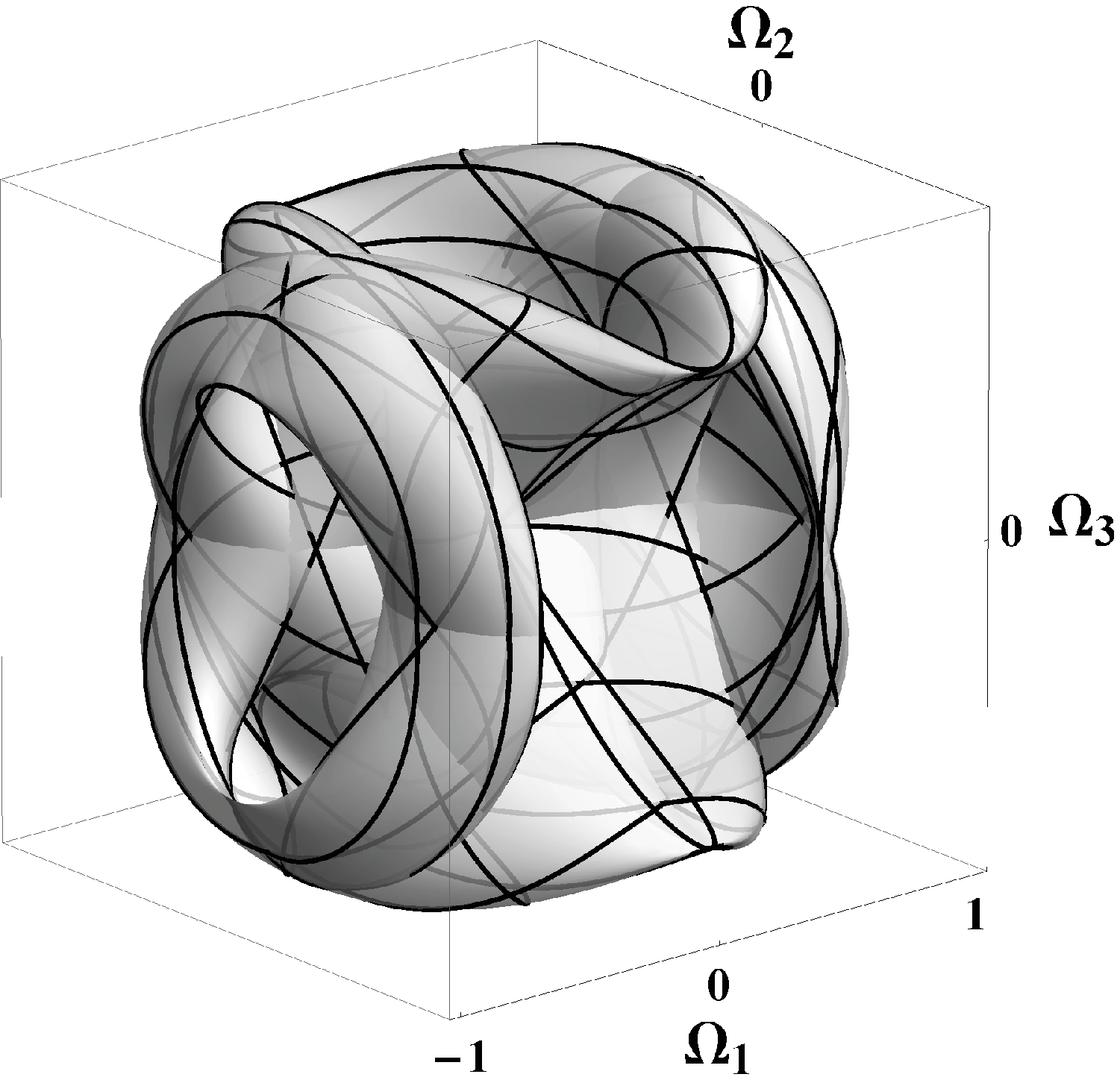}\end{tabular}
\\
\hline

\begin{tabular}{c}\ts{II}\end{tabular} &
\begin{tabular}{c}\includegraphics[width=\TableTwoCof\linewidth,keepaspectratio]{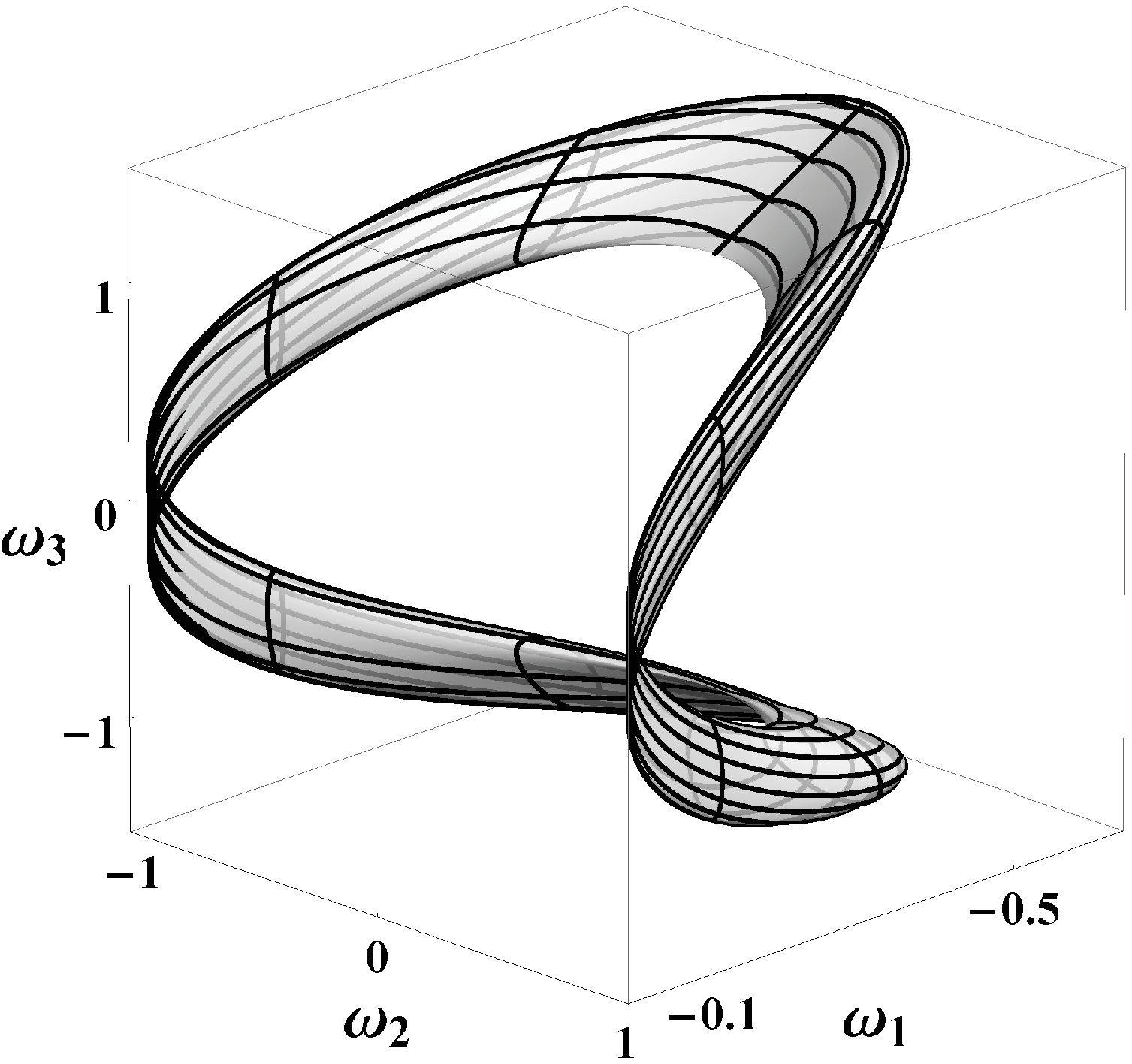}\end{tabular} & \begin{tabular}{c}\includegraphics[width=\TableTwoCof\linewidth,keepaspectratio]{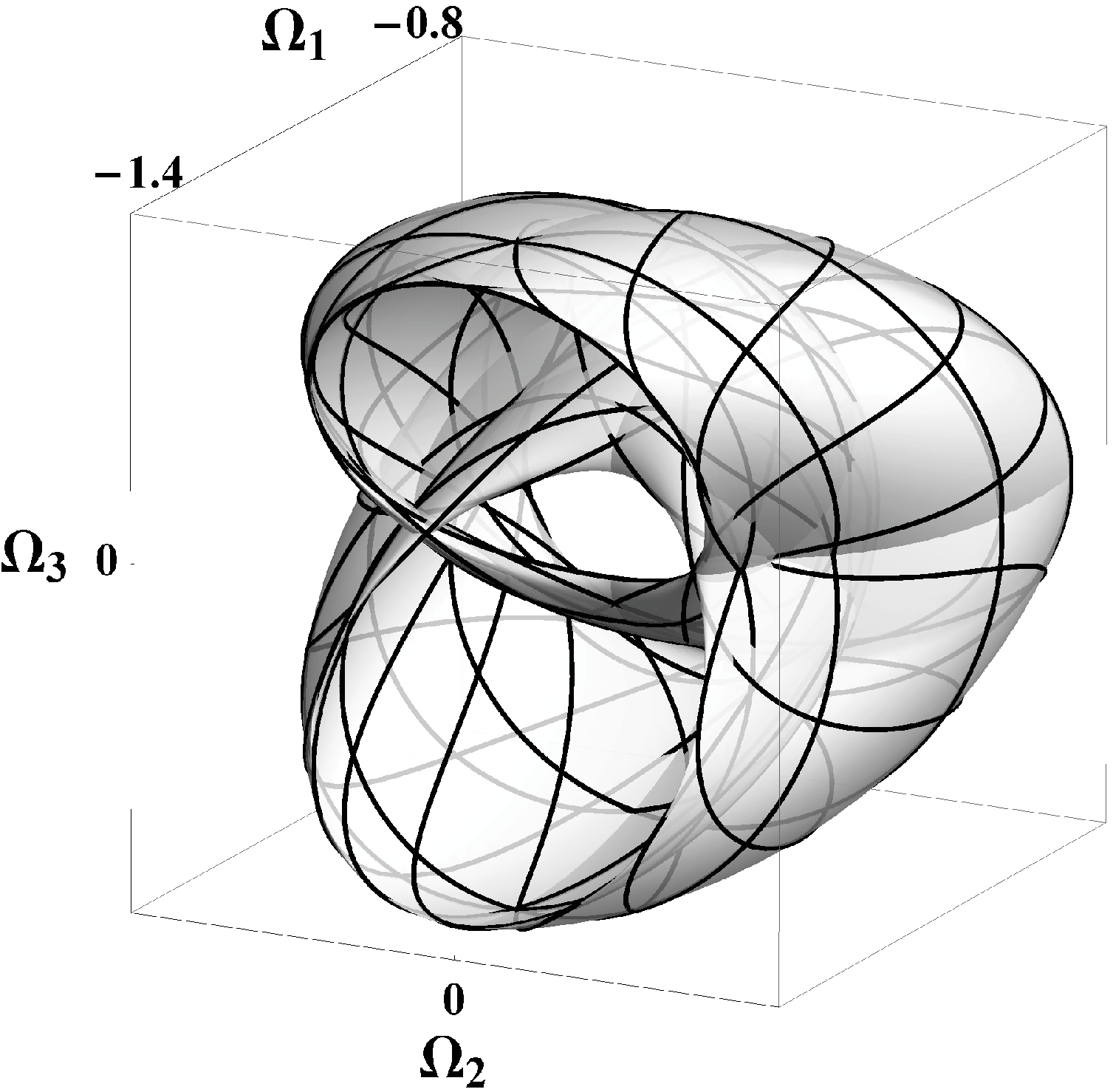}\end{tabular}\\
\hline
\begin{tabular}{c}\ts{III}\end{tabular} &
\begin{tabular}{c}\includegraphics[width=\TableTwoCof\linewidth,keepaspectratio]{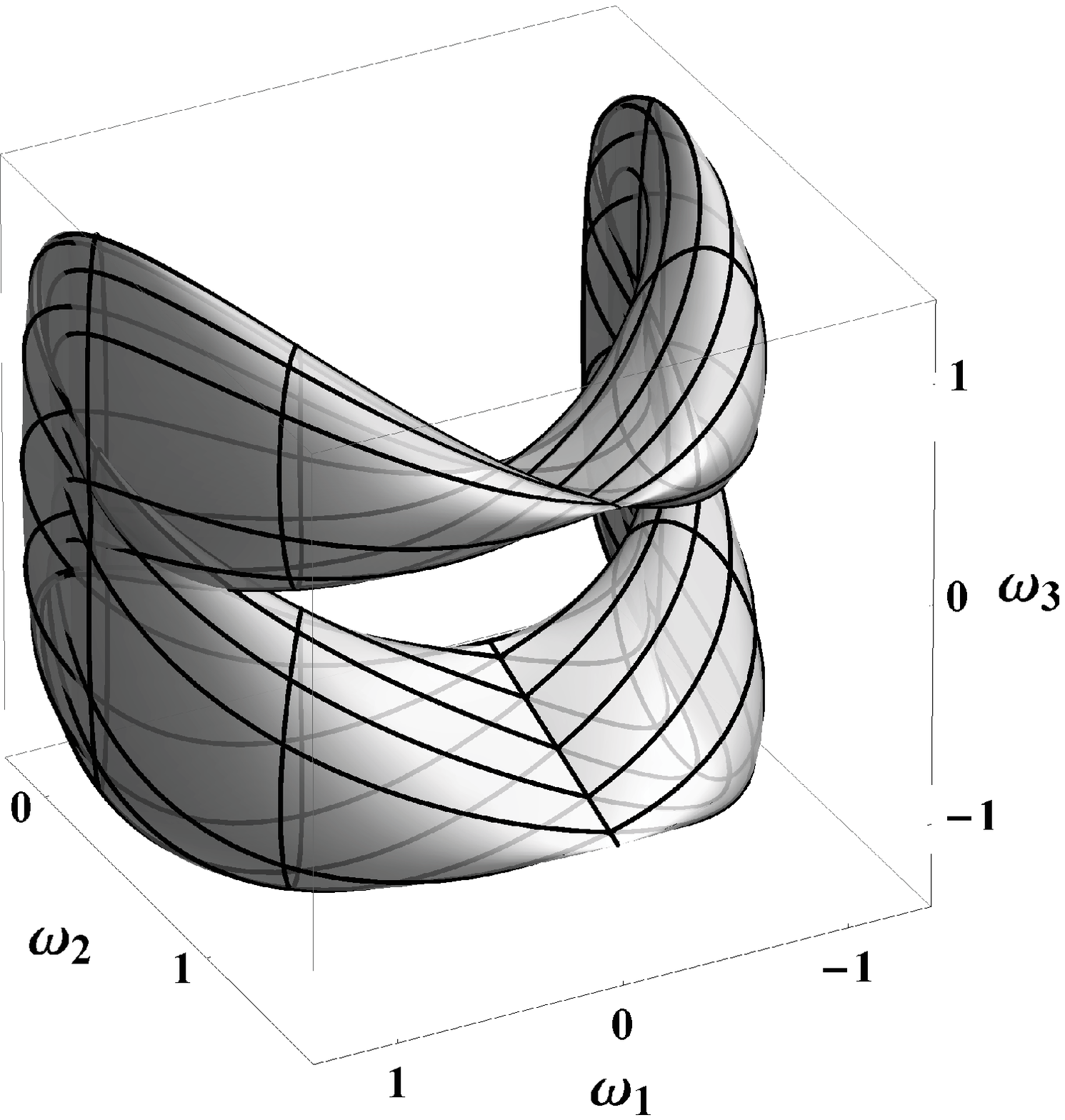}\end{tabular} & \begin{tabular}{c}\includegraphics[width=\TableTwoCof\linewidth,keepaspectratio]{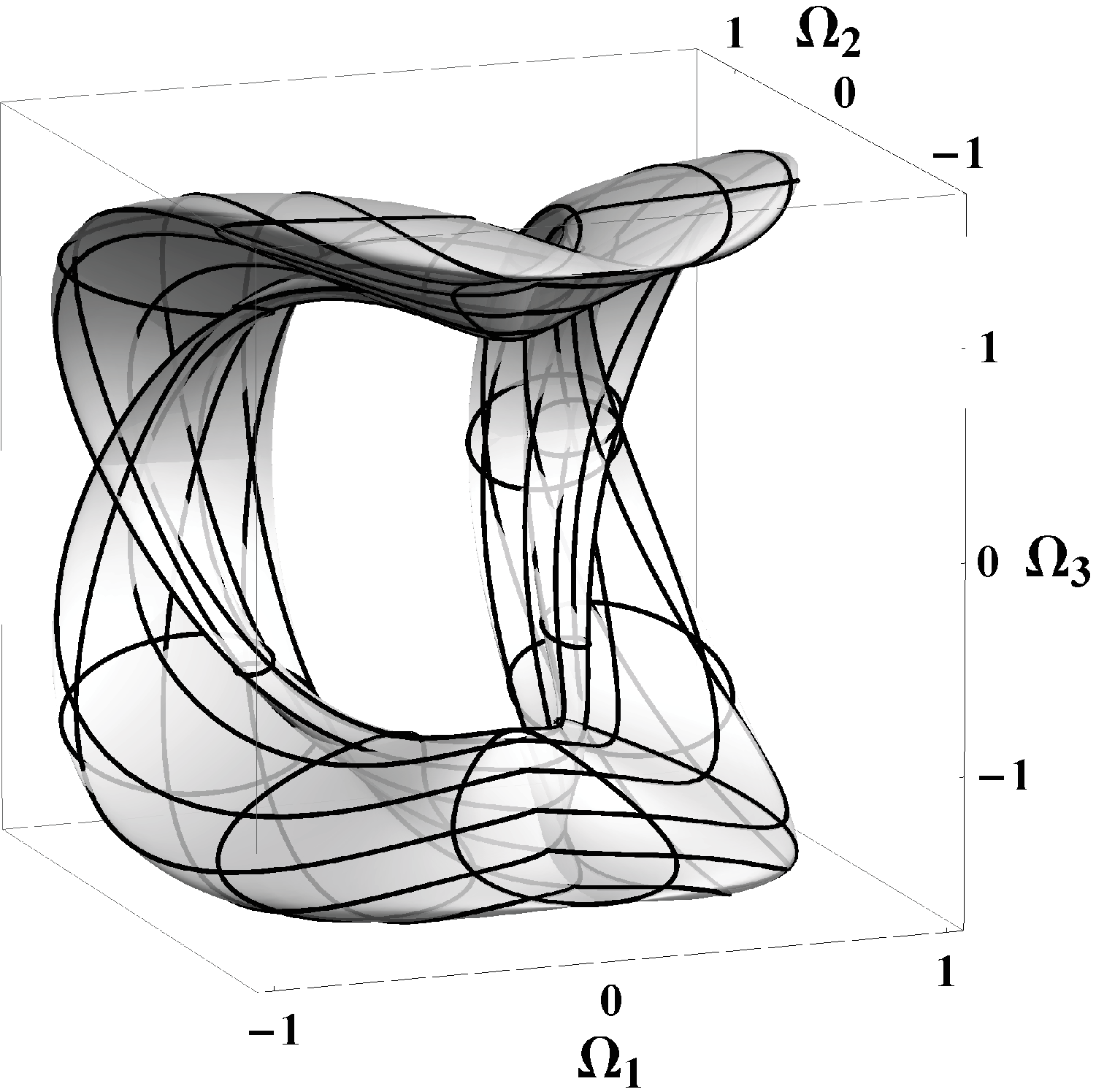}\end{tabular}\\
\hline

\begin{tabular}{c}\ts{IV}\end{tabular} &
\begin{tabular}{c}\includegraphics[width=\TableTwoCof\linewidth,keepaspectratio]{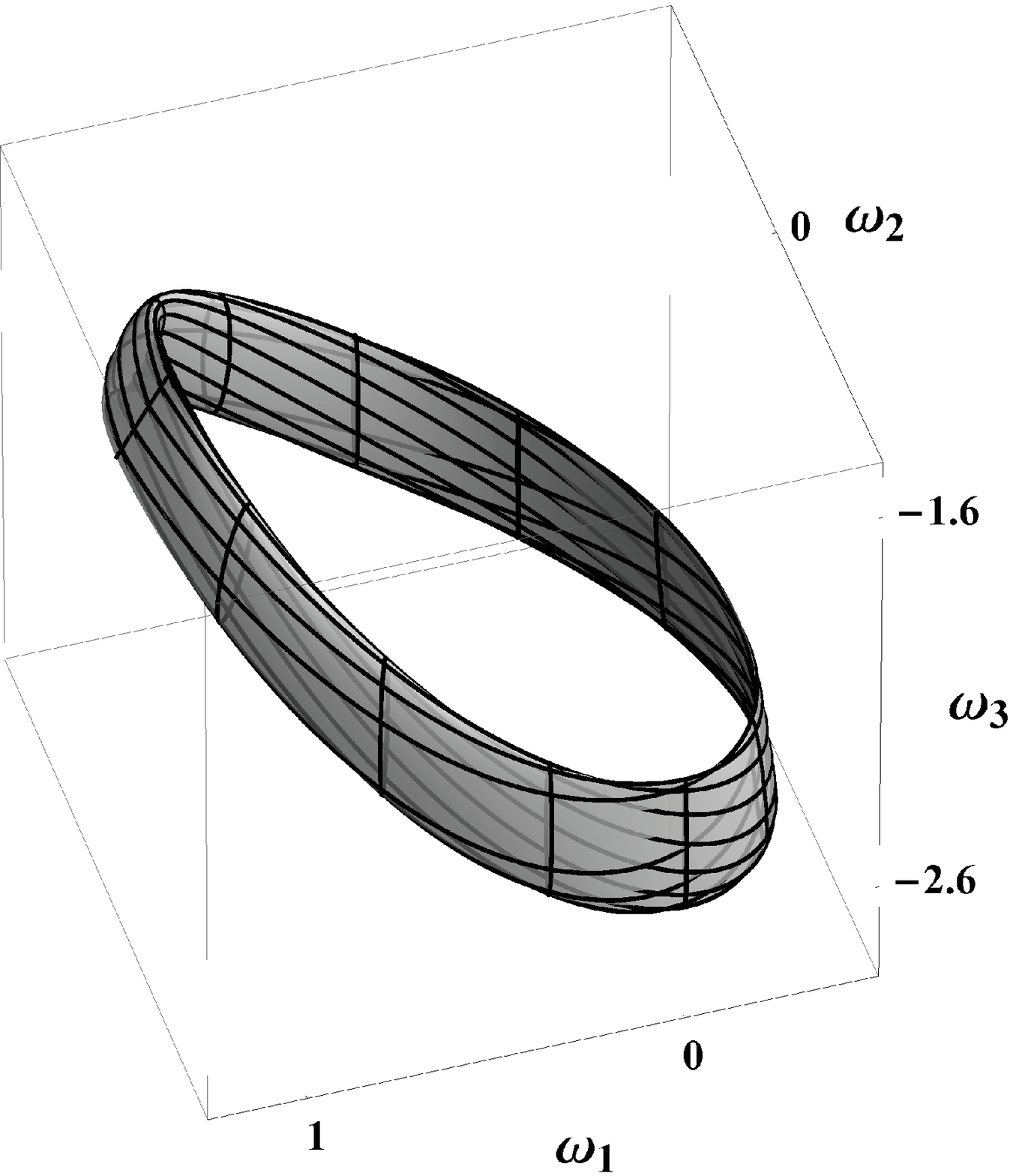}\end{tabular} & \begin{tabular}{c}\includegraphics[width=\TableTwoCof\linewidth,keepaspectratio]{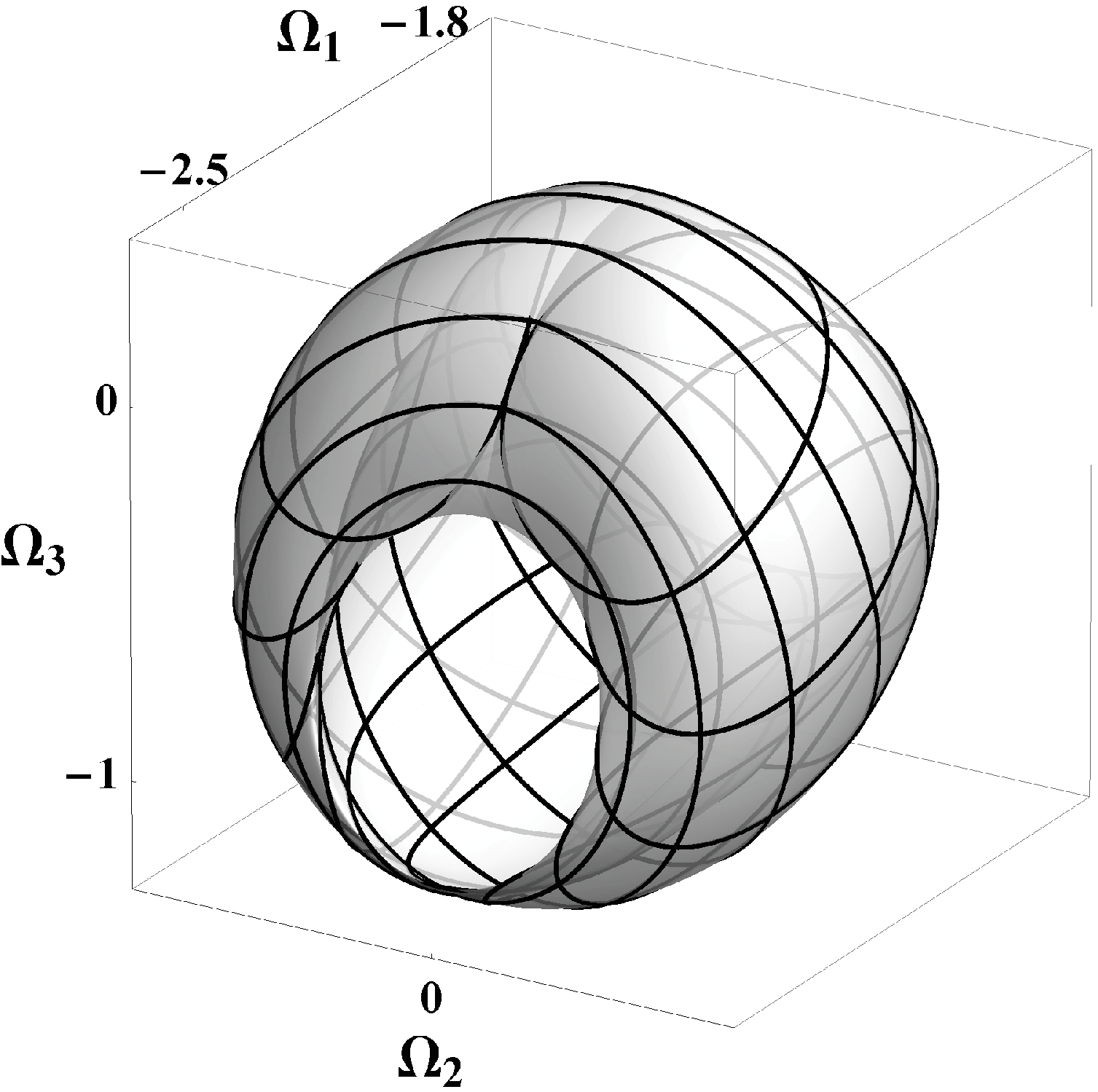}\end{tabular}\\
\hline
\begin{tabular}{c}\ts{V}\end{tabular} &
\begin{tabular}{c}\includegraphics[width=\TableTwoCof\linewidth,keepaspectratio]{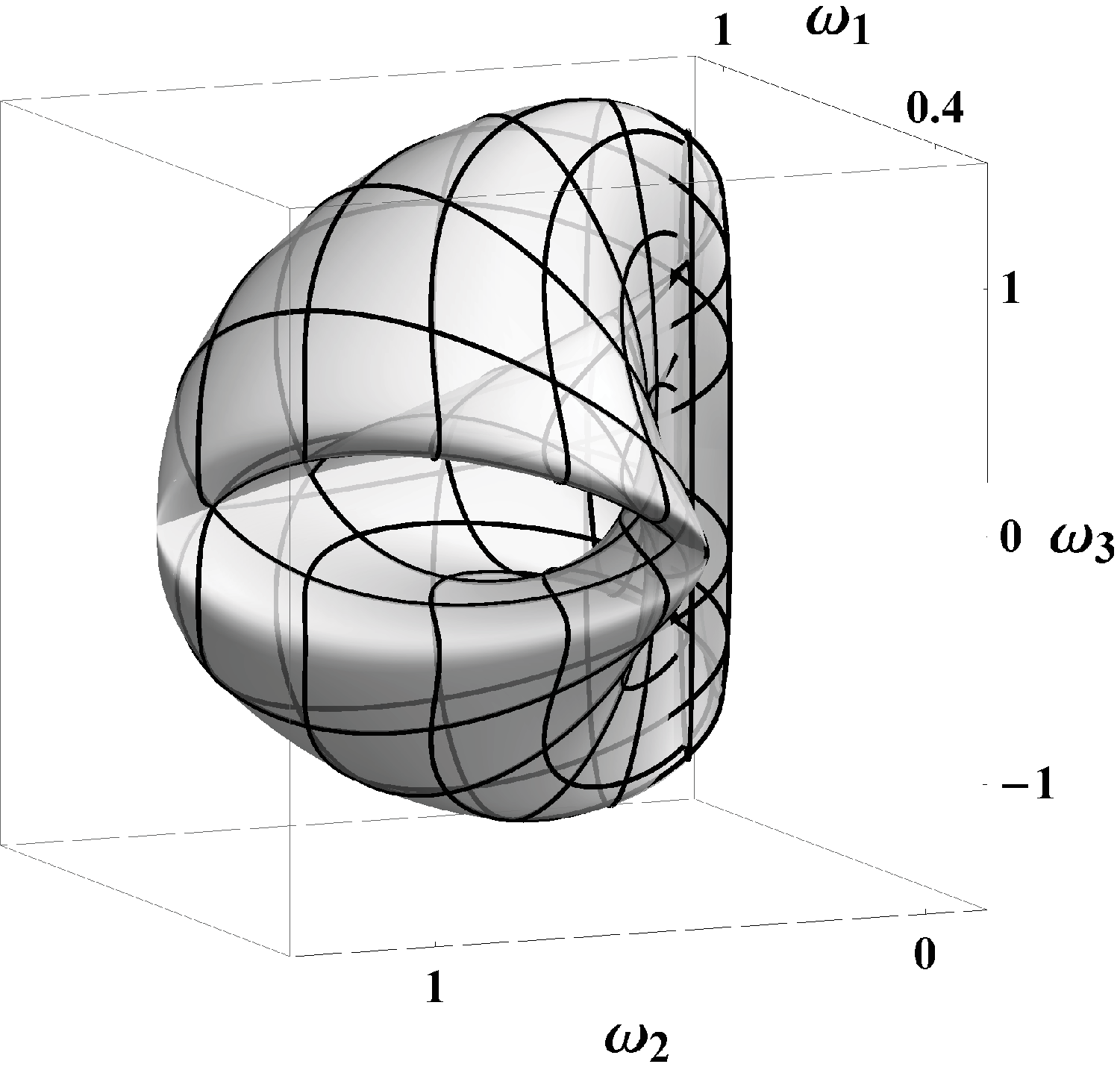}\end{tabular} & \begin{tabular}{c}\includegraphics[width=\TableTwoCof\linewidth,keepaspectratio]{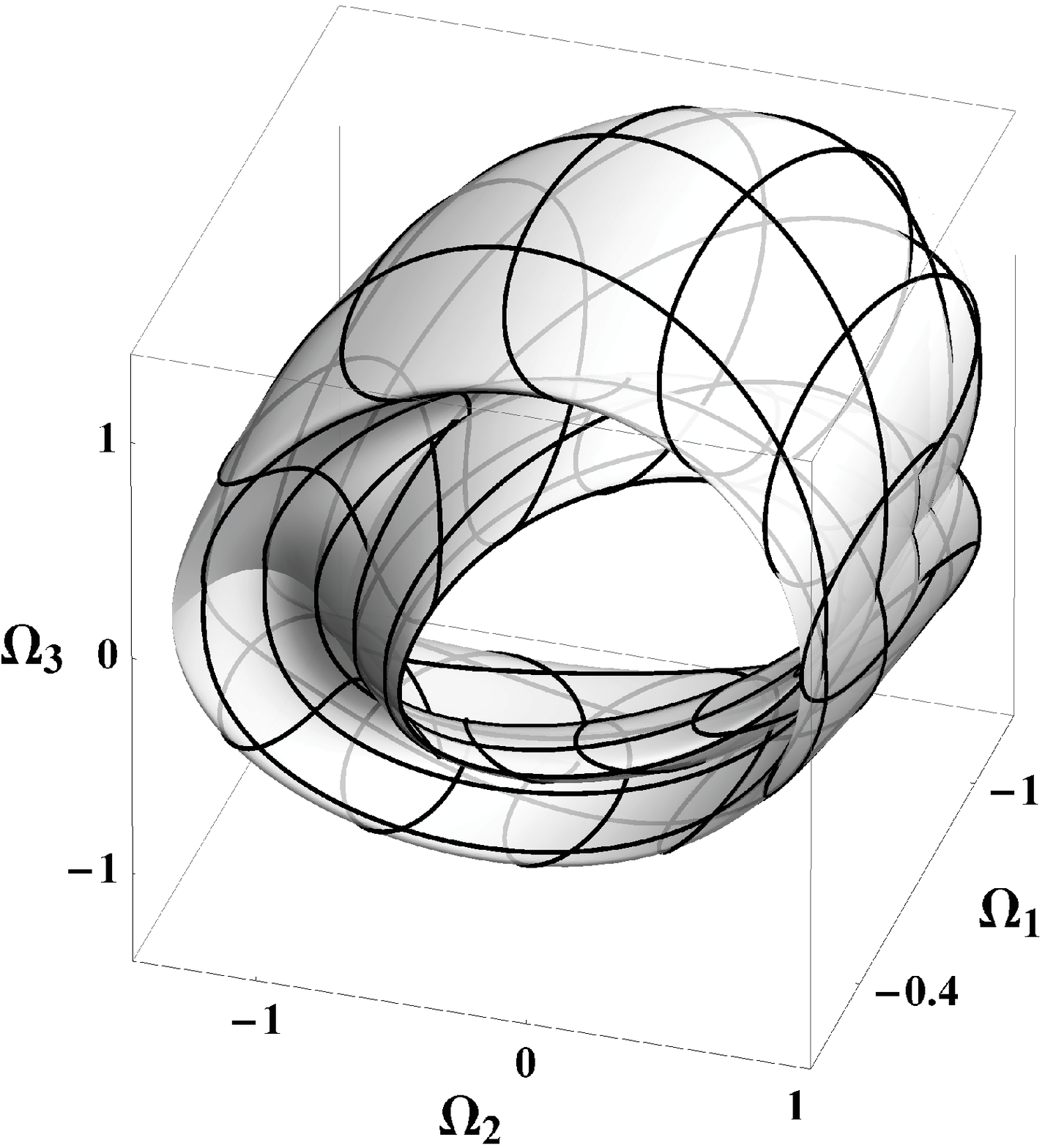}\end{tabular}\\
\hline
\begin{tabular}{c}\ts{VI}\end{tabular} &
\begin{tabular}{c}\includegraphics[width=\TableTwoCof\linewidth,keepaspectratio]{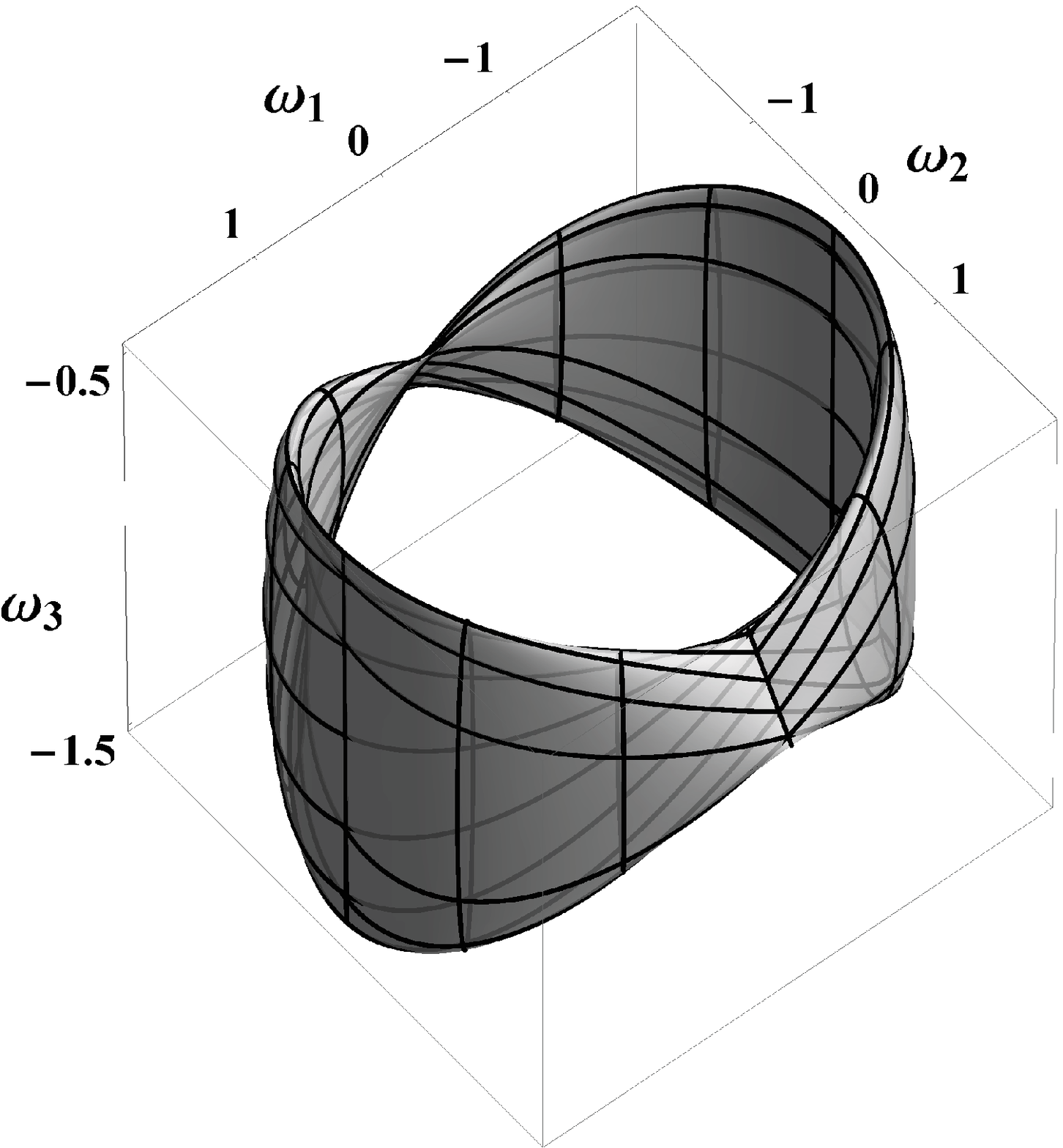}\end{tabular} & \begin{tabular}{c}\includegraphics[width=\TableTwoCof\linewidth,keepaspectratio]{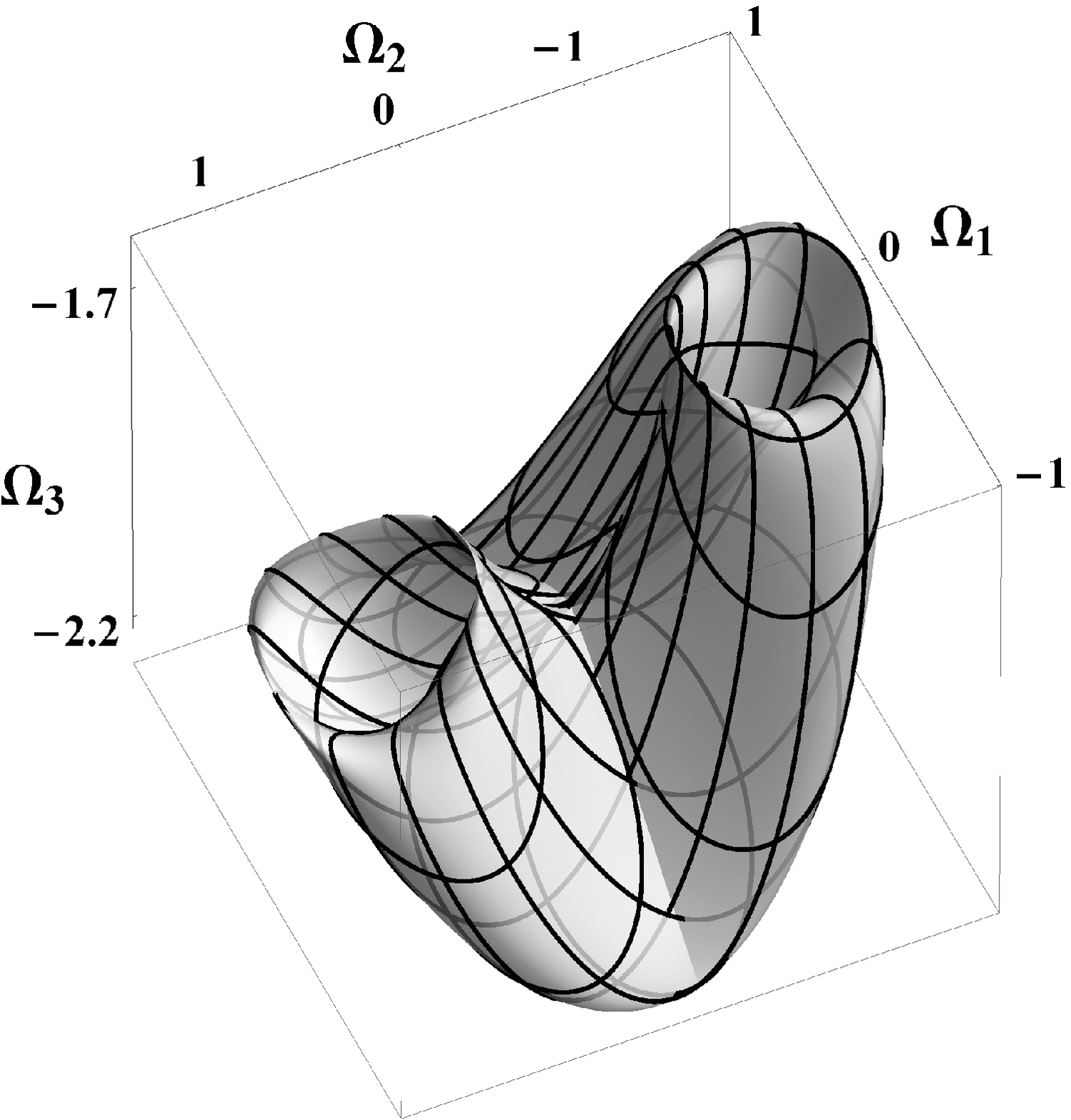}\end{tabular}\\ \hline
\end{longtable}

The result of the computer visualization of the surfaces bearing the moving and fixed hodographs is shown in Table~2. Using the explicit formulas for the orientation matrix we obtain the geometric representation of motion as rolling without slipping of a quasi-periodic curve on the $P$-surface over a similar curve on the $\Pi$-surface.

Let us use the obtained above decomposition of motions and point out some properties of the arising pendulum type motions. To this end, we write $\bo'$, $\bo''_{(1)}$ and $\bo''_{(2)}$ from  \eqref{eq2_6}, \eqref{eq2_8} substituting \eqref{eq3_2}. Then we have
\begin{equation}\notag
\begin{array}{c}
  \displaystyle \bo' = \left(0,0,-\frac{\SFR _1 \SR _2 + \FR_2 \SSR _1}{2(1-s_1^{-1}s_2)}\right), \\
  \displaystyle \bo''_{(1)} = -\frac{\SFR _1}{2(1- s_1^{-1} s_2)} \left(  0, - r, \SR _2  \right), \qquad
  \displaystyle \bo''_{(2)} = -\frac{\FR _2}{2(1- s_1^{-1}s_2)} \left(  s_1^{-1} r , 0, \SSR _1  \right).
\end{array}
\end{equation}
It follows from Table~1 that $\SFR_1$ periodically changes its sign in the cameras $\ts{I},\ts{II},\ts{IV}$ and has constant sign in $\ts{III},\ts{V},\ts{VI}$. The value of $\FR_2$ periodically changes its sign in the cameras  $\ts{I},\ts{III},\ts{VI}$ and has constant sign in  $\ts{II},\ts{IV},\ts{V}$. The value of $\SSR_1$ changes its sign in all cameras except for $\ts{IV}$, and $\SR_2$ changes its sign in all cameras except for $\ts{VI}$.
In particular, in all cameras the products $\SFR_1 \SR_2$ and $\FR_2 \SSR_1$ change the sign periodically. This means that, for the semi-moving frame its pendulum motion is always an \textit{oscillation}.
The impact on this motion of the separation variables can be estimated using \eqref{eq2_7}.

The motions of the body in the semi-moving frame are superpositions of two partial motions; each partial motion is of pendulum type. The first one is defined by the evolution of the radical $\SFR_1$ and is an \textit{oscillation} in the cameras $\ts{I},\ts{II},\ts{IV}$ and a \textit{rotation} in the cameras $\ts{III},\ts{V},\ts{VI}$. The second partial motion according to the evolution of the radical $\FR_2$ is an \textit{oscillation} in the cameras
$\ts{I},\ts{III},\ts{VI}$ and a \textit{rotation} in the cameras $\ts{II},\ts{IV},\ts{V}$. These properties also explain the position of the bearing surfaces shown in Table~2 with respect to the coordinate planes.

\section*{Acknowledgements.} The work is partially supported by RFBR, research project No.\,13-01-97025.


\begin{thebibliography}{10}

\bibitem{Po1851}
L.~Poinsot, Th\'{e}orie nouvelle de la rotation des corps, J. de Math. Pures et
  Appl. 1~(16) (1851) 289--336.

\bibitem{Kh1964}
P.~V. Kharlamov, Kinematic interpretation of the motion of a body with a fixed
  point, Journal of Applied Mathematics and Mechanics 28~(3) (1964) 615--621.
\newblock \href {http://dx.doi.org/10.1016/0021-8928(64)90102-9}
  {\path{doi:10.1016/0021-8928(64)90102-9}}.

\bibitem{GorrMono}
G.~V. Gorr, L.~V. Kudryashova, L.~A. Stepanova, Classical problems in the rigid
  body dynamics. \mbox{T}heir developement and current state, Kiev: Naukova
  dumka, 1978.

\bibitem{KhKh1983}
M.~P. Kharlamov, P.~V. Kharlamov, To solve a problem of rigid body dynamics.
  {W}hat does it mean, in: Proc. of the IUTAM-ISIMM Symp. on Modern
  Developments in Analytical Mechanics, Vol.~2, Accademia delle Scienze di
  Torino, Torino, 1983, pp. 535--562.

\bibitem{GaGoKov}
I.~N. Gashenenko, G.~V. Gorr, A.~M. Kovalev, Classical problems of the rigid
  body dynamics, Kiev: Naukova Dumka, 2012.

\bibitem{Ga1986}
I.~N. Gashenenko, Moving hodograph of the angular velocity in the solution of
  \mbox{G}oryachev--\mbox{C}haplygin, Mekh. Tverd. Tela 18 (1986) 3--9.

\bibitem{Ga1988}
I.~N. Gashenenko, Fixed hodograph of the angular velocity in the solution of
  \mbox{G}oryachev--\mbox{C}haplygin, Mekh. Tverd. Tela 20 (1988) 29--34.

\bibitem{Ga1990}
I.~N. Gashenenko, Geometrical analysis of two-frequency quasi-periodic motions
  of the \mbox{K}ovalevskaya gyroscope, Mekh. Tverd. Tela 22 (1990) 1--10.

\bibitem{KhRCD}
M.~P. Kharlamov, Bifurcation diagrams of the \mbox{K}owalevski top in two
  constant fields, Regular and Chaotic Dynamics 10~(4) (2005) 381--398.
\newblock \href {http://arxiv.org/abs/0803.0893} {\path{arXiv:0803.0893}},
  \href {http://dx.doi.org/10.1070/RD2005v010n04ABEH000321}
  {\path{doi:10.1070/RD2005v010n04ABEH000321}}.

\bibitem{Odin}
M.~P. Kharlamov, One class of solutions with two invariant relations of the
  problem of motion of the \mbox{K}owalevski top in double constant field,
  Mekh. Tverd. Tela 32 (2002) 32--38.
\newblock \href {http://arxiv.org/abs/0803.1028} {\path{arXiv:0803.1028}}.

\bibitem{KhSavUMBeng}
M.~P. Kharlamov, A.~Y. Savushkin, Separation of variables and integral
  manifolds in one problem of motion of generalized \mbox{K}owalevski top,
  Ukrainian Mathematical Bulletin 1~(4) (2004) 569--586.
\newblock \href {http://arxiv.org/abs/0803.0882} {\path{arXiv:0803.0882}}.

\bibitem{Yeh1}
H.~M. Yehia, New integrable cases in the dynamics of rigid bodies, Mechanics
  Research Communications 13~(3) (1986) 169--172.

\bibitem{ReySem}
A.~G. Reyman, M.~A. Semenov-Tian-Shansky, Lax representation with a spectral
  parameter for the \mbox{K}owalewski top and its generalizations, Lett. Math.
  Phys. 14~(1) (1987) 55--61.

\bibitem{SavPHD}
A.~Y. Savushkin, Investigation of one class of exact solutions in the problem
  of motion of the \mbox{K}ovalevskaya top in a double force field, Ph.D.
  thesis, School: Volgograd, M.P.Kharlamov, Volg. Acad. of Publ. Admin. (2004).

\bibitem{KhND11}
M.~P. Kharlamov, Topological analysis and \mbox{B}oolean functions. \mbox{II}.
  \mbox{A}p\-p\-lication to new algebraic solutions, Nonlinear Dynamics 7~(1)
  (2011) 25--51.
\newblock \href {http://arxiv.org/abs/1309.7180} {\path{arXiv:1309.7180}}.

\end{thebibliography}
\end{document}